# The Role of PMESII Modeling in a Continuous Cycle of Anticipation and Action

Alexander Kott and Stephen Morse

The inevitable incompleteness of any collection of PMESII models, along with poorly understood methods for combining heterogeneous models, leads to major uncertainty regarding the reliability of computational tools. This uncertainty is further exacerbated by difficulties in validation of such tools. Throughout this book, we stressed this uncertainty repeatedly, and urged our readers to use such tools with great caution, indicating that these tools should only be used as aids to human analysis and decision-making.

We believe that prudent, problem-specific methodologies are indispensable when using computational tools. Strictly speaking, it would be outside the scope of this book to outline such methodologies when using the modeling and estimating tools based on computational techniques discussed here.

However, given the practical orientation of the book, we cannot ignore entirely the question of how practitioners can make better use of computational tools. In particular, a practitioner must wonder: how can we accommodate the uncertainty of a tool's results by applying human judgment appropriately?

## 1. Uses of Estimates

In two examples we discuss below, planners and analysts used (or could have used) computational tools to obtain estimates of effects of various actions under consideration. Then they considered these computational estimates to draw their own conclusions regarding the effects that would likely emerge from proposed actions taken by the international mission. The specific ways in which they could utilize these estimates deserve elaboration. Here we rely on observations made during extensive experiments with a suite of PMESII tools (Kott 2007)

First, the most conventional use of computational estimates occurs when the user largely agrees with the estimates. She finds the estimates consistent with her intuition and expectations. In this case, the main value of computational tools is in producing a far greater degree of detail than a human analyst or expert can provide. The estimates of a PMESII tool can describe, for example, the changes in hundreds of diverse political, economic, and social variables over many years, with a time-step of one week. Even a large team of expert analysts cannot produce such a detailed product (assuming a decision requires this depth of analysis) in a practical amount of time. Still, the user may find the need to adjust some of the estimates, or replace part of them with her expert judgments.

A second case occurs when the user originally disagrees with the estimates. They do not appear reasonable and contradict her intuition and experience. However, having reviewed the estimates in detail and having traced the dependencies between the elements of the estimates, the user often reconsiders her views. She finds that the computational estimates included factors she has not contemplated. Upon further reflection, she favors the computational estimates. Nevertheless, like in the first case, the user may choose to modify some of the estimates.

Third, even if the user disagrees with the estimates, the model can still point out potential effects that the user has not appreciated beforehand. Consider that a PMESII tool often produces estimates that describe changes in hundreds of variables. Even a highly experienced analyst is unlikely to consider all of such variables. Having examined the estimates, however, the user often finds an important effect—a change in a PMESII variable—she has not considered and now finds important to investigate further, even if she disagrees firmly with the direction and magnitude of the computer-estimated effect.

Fourth, a computational tool is an important mechanism to elucidate and examine assumptions, which otherwise remain hidden or untested.  As the user explores or explains the reasons she disagrees with the computer-generated estimates, she often verbalizes heretofore unspoken assumptions that should have been recorded or perhaps revisited. Occasionally, the user herself is unaware of implicit assumptions, which the process of interacting with a computer tool helps to reveal. When this learning process occurs among a team of analysts, one may be surprised by the assumptions made by other experts. Thus, computer-generated estimates serve as a helpful learning mechanism to draw out unrecognized assumptions and biases of human experts.

Fifth, in a similar fashion, a computer tool helps the user by providing an alternative or opposing opinion. By examining the computational results and formulating the reasons why she disagrees with them, the user arrives to a clearer, more logical and compelling explanation of her own position. In effect, the computational models serves here as a useful intellectual punching bag with which the user fortifies her own arguments.

Fifth, in a similar fashion, a computer tool helps the user by providing alternative, sometimes conflicting views. By examining the computational results and formulating the reasons why she disagrees with them, the user arrives at a clearer, more logical and compelling explanation of her own thinking. In effect, a computational model serves here as a useful intellectual punching bag with which the user learns and informs her own judgments.

## 2. Analogy:  Model Predictive Control

We should also ask a broader question: how should one choose an action (or decide to do nothing) in an uncertain world when the computational means for anticipating the action's impact are imperfect?

This very question is a subject of rigorous and extensive research in the discipline of control theory. Some key challenges and solutions developed in that discipline can give us useful insights in answering our broader question mentioned above.

A simple example of a control system is described in Figure 1. On the left side is a depiction of an aircraft control system. An automatic system controller continuously senses the behavior of the aircraft (the "plant", in control-theoretic parlance); and based on an analysis of the sensor readings, it then issues control signals to the aircraft to produce the desired behavior. For an airplane, for example, the sensors tell the controller about such things as air speed, pressure, and wind; and then the controller issues signals that automatically adjust the flaps and power to keep the plane on course (the course having been selected by the pilot).

INSERT Figure 1. An analogy between an aircraft control system and an intervention management system

In the context of an international invention shown on the right side of Figure 1, the "plant" is the troubled country experiencing a crisis and in need of international assistance. The control signals are the directed actions issued by the intervention managers (let us call them so for the sake of generality) of the international mission regarding actions to be taken on the ground. These might include directions for enforcing a cease-fire, providing food aid, or arresting leaders of a criminal organization. Instead of sensor information in an automatic controller of an aircraft, intervention managers make observations and get reports of the situation within the country, such as reports of unwanted firefights, trends of malnutrition, or incidents perpetrated by a criminal organization.

Just as an automatic control system *continuously* monitors the state of the aircraft in flight and adjusts accordingly, so also should the intervention managers perform a periodic review of progress achieved so as to flexibly exploit opportunities and rapidly respond to unwanted events within the country.

In the discipline of control theory, it is typically assumed that a *model of the system* to be controlled is available. The model can relate controllable inputs and the state of the current system so as to determine likely outcomes – that is, forecast how the current state of the system would change in response to a variety of user-selectable actions. Given a set of desired outcomes – to keep an aircraft stable and on course, for example – the controller uses the outcomes derived from the model to select from among available actions those which would best advance the desired goals.

But what if the model is imperfect? How would an intervention manager choose the most effective actions to make things better if the available model could not accurately represent the actual behavior of the society?

Control theorists have studied how the controller should behave if there is a disparity between the system model, on the one hand, and the actual system state and behavior, on the other. That is, if the model predicts one type of response, but the system will in fact behave in some other way, the potential exists for erroneous (and perhaps catastrophic) outcomes to emerge. A control system should be able to perform within specified constraints, even when certain types of model disparities are present.

One method that control theory provides foe such complex control problems is *model predictive control* (MPC). It is particularly appropriate when we are concerned about significant disparities between our model and the actual "plant," as is inevitably the case when modeling an intervention in a country in crisis.

The key steps of the MPC approach are to (a) perform predictive modeling of the plant for a relatively long period of time from the current moment forward; (b) select control signals that optimize long-term performance of the plant; (c) after a short period of executing the control signals, repeat the process of modeling (starting with an assessment of the new, most recent observations of the current situation); and (d) adjust the control signals (e.g., Mayne 2000). In this manner, by frequent re-modeling followed by timely adjustment of the control signals, the approach of MPC reduces the erroneous effects associated with the model's limitations in depicting reality in an uncertain world.

Let us elaborate on the MPC approach using the terminology associated with managing an international intervention. At a suitable time point during an intervention, the intervention managers assess reports on the current situation, reconsider their current strategy and then revisit their plan of action, which covers a period of time called the planning horizon, let's say 18 months forward. Then, they use the available PMESII models to assess the effects of their plan for the duration of the planning horizon. They will probably repeat the planning-modeling process several times, until they arrive to the plan that seems to promise the best combination of effects. Having generated a revised plan of action, they issue guidance and orders to various field organizations that carry out the actions directed by the intervention managers.

As time passes and field organizations perform the specified actions, the intervention managers collect further observations regarding progress achieved. When they detect an unexpected or unwelcome event—either a major crisis, or a serious deviation from the anticipated effects, or simply a pre-defined number of months pass—the intervention managers perform replanning (Figure 2). Again, they adjust their plan based on the most recent situation reports, use the available PMESII models to estimate impacts 18 months forward, adjust their current plan to produce the best combination of effects from this time forward, and issue new guidance and orders to the field organizations.

Note that the assess-remodel-adjust plan cycle tends unfold within significantly shorter time period than the 18 month panning horizon. The planning horizon could be, for example, equal to six update cycles; that is, the plan is revised six times (say, every 3 months) before its horizon (18 months) elapses. For this to happen, the international mission should be organized and staffed to perform such recurring updates. Here, periodic control activities (that is, frequent assess-remodel-adjust plan) are not the exception, but rather they are the expected, normal approach for managing an international intervention in complex and uncertain environments.

INSERT Figure. Continuous replanning process is analogous to the model predictive control

To be sure, the model predictive control concept cannot be applied to intervention management in a blind, mechanistic fashion. Human beings are not mere sensors and actuators, and real-world interventions are immensely difficult to manage, often involve tragic upheavals that require policy-makers and decision-makers to apply deep human insights, judgment, experience, and leadership. Indeed, there are many complications.

Frequently, it is difficult to collect and interpret observations and reports regarding the country's situation at any given moment. Information is often incomplete, partly erroneous or intentionally distorted, contradictory, and subject to conflicting assessments. Deciding on the true meaning of the available information requires experience and mature judgment. Here, however, use of computational models can help highlight most consistent interpretations.

There are many real world examples. Graham-Brown (1999) provides a cautionary note. In describing the experiences of international sanctions imposed on Iraq beginning in 1991, she illustrates the extreme difficulties that intervening organizations—governmental or non-governmental—find in collecting reliable information on effects of intervention actions, and the multiple distortions of that information. One must not underestimate the challenges of objectively assessing conditions on the ground.

Further, as we already emphasized, the PMESII models are inevitably imperfect in modeling a country in crisis and they can produce results which can be potentially misleading in ways that may be unknown to users. Although MPC paradigm serves to reduce the impact of such imperfections, users of models must apply common sense and interpret the computational results critically. When significant disparities occur between the model estimates and real-world results, the models may need adjustments and even major modifications.

The key point is that a mission's plan of action in an uncertain world is not immutable, but only reflects current best knowledge. Consequently, it should be updated as new information on the situation becomes available. In an uncertain world, the MPC approach of periodic assessment and a rolling time-horizon offers a structured way of organizing reporting systems, revisiting current strategies for change, and adjusting the mission's plan of action. This approach should be tailored to the specific requirements and constraints of a particular task and organization. We now describe two examples where a conceptually similar approach to MPC is applicable.

## 3. Example: Advance Planning

In 2009, a US government organization completed a detailed study that generated an advance plan for assisting a certain friendly country (referred to here as "Country-Y") that was experiencing a threat of insurgency. In performing this planning effort, the organizers of the study used a collection of PMESII models and developed a methodology that bears partial resemblance to the MPC paradigm. The following discussion uses only the publically available information reported in (Messer, 2009).

The purpose of the study effort was to develop a preliminary, high-level intervention plan that involved assisting the government of Country-Y in defending itself against the insurgency while building an indigenous security capacity. The product of the study effort would provide an analytical baseline for future additional planning, if any were to become necessary.

The study included a broad range of experts from several US government agencies and one non-US government organization. These experts represented many diverse areas: agriculture, justice, commerce, diplomacy and state relations, international development, international finance, intelligence gathering, and military services.

The study made use of numerous PMESII tools including several that we discussed in this book: COMPOEX, NEXUS, PSOM, Senturion, and others. To combine the use of PMESII models with expert judgments, the organizers of the study developed a human-in-the-loop wargaming procedure they called the X-Game, which resembles the MPC approach in several aspects. Like a conventional wargame, the X-Game involved several teams of human analysts, planners, and subject-matter experts.

Participants were organized into five cells. A Blue Cell played the role of the intervening nations and officials of the government of Country-Y. This cell developed plans of actions intended to accomplish the objectives of the international intervention. A Red Cell played the role of various opposition parties, such as internal insurgent group, that threatened the government of Country-Y as well as the international mission. This cell defined objectives and plans of actions of all such opposing parties. A Green Cell played the role of various groups within the indigenous population and reflected their changing attitudes and reactions. The White Cell used subject-matter experts to assess the effects of all the actions on a broad range of PMESII-type metrics. In addition, a Modeling Team operated the PMESII computational models.

The overall planning horizon was ten years. This long period was divided into phases. In the each phase, the following process took place, approximately:

1. Taking into account all available information about the situation in Country-Y at the beginning of this phase, and the history of events up to that point, the Blue, Red and Green Cells formulated their plans of actions several years forward. They presented these plans to the Modeling Team and the White Cell.

2. The Modeling Team entered the plans into PMESII modeling tools and used the tools to generate estimates of PMESII effects as they unfolded to the end of the entire 10-year period. This step was also a good time to adjust the models if there was a major difference between expert assessments of effects of the previous phase and the corresponding computational estimates.

3. The Modeling Team presented the computed results to the White Cell.

4. The White Cell generated the assessment of how PMESII effects would unfold in Country-Y over time, taking into account the Blue and Red actions. In this assessment, the White Cell considered the predictions of the computational tools, but did not necessarily follow these predictions.

5. The White Cell decided when a significant change occurred in the situation of Country-Y. This point in time became the beginning of the next phase, and the process iterated.

The organizers of the study reported achieving the required objectives of the study effort. To our knowledge, this was the first large-scale study to make systematic use of multiple PMESII computational tools for a practical planning purpose.

## 4. Example: Next State Planning

The second example is decribed by Kott et al (2007a) in what is now called the approach of Next State Planning for organizing planning efforts of an international mission. They point out that use of PMESII computational tools would fit naturally into the approach for Next State Planning.

The international intervention in Kosovo in 1999 planted the seeds of Next State Planning. The combined UN–NATO mission (called UNMIK-KFOR) initiated this approach by preparing short-term plans to achieve specific near-term objectives on the ground in the coming 3-5 months. This near-term approach was taken in addition to developing the mission's overall plan under a much longer time horizon of three years, so called End State Planning.

The method of Next State Planning derives from the lessons learned in the Kosovo intervention, and generalizes some of the planning methods actually used there (Covey et al 2005). Although mission managers in Kosovo did not use PMESII modeling tools discussed here, they do identify a beneficial role such tools could play if they were available at the time. Interestingly, while different from the X-Game, the Next State Planning approach also bears resemblance to the MPC paradigm in certain important aspects.

Although much smaller in size than later interventions, the UN-NATO intervention in Kosovo was no less extensive or complex. The UN-led civilian mission included 5,000 international police and 3,000 civil administrators, advisors, and trainers. Many humanitarian non-governmental organizations sent thousands of relief workers and human rights investigators. Moreover, the NATO-led military command, KFOR, numbered 44,000 troops.

Based on the Kosovo experience, Covey et al (2005) argue that planning of an intervention should occur on more than one temporal scale. At the most extensive, a three-year mission plan for the entire intervention is needed to envision the emergence of the lasting political solution to the conflict. However, this long-range mission plan cannot attach great certainty to many key developments in the near term anticipated to a successful intervention, and, of necessity, defines a range of alternative approaches rather than a specific path. At this highest level of abstraction, the challenge is in ensuring that at least one of the paths will prove viable and effective once details are understood.

Figure 3 illustrates how an international mission, operating in the presence of uncertainty, engages in a continuous process of moving in the direction of the desired end-state, but in a step-wise process of achieving a series of "next states". This "next state thinking" involves continuously updating the mission's understanding of the situation and projecting it forward in time, and using this projection to guide near-term planning to achieve a desired "next state". The overall result is a jagged path of desired next states with a time horizon of 3-5 months that moves forward (seemingly haphazardly) but relentlessly in the direction of the desired end-state over a longer time horizon of 3-5 years.

INSERT Figure 3: High-level intervention plan faces a wide range of uncertainties

Consequently, to ensure that the path proceeds in the direction of the overall desired end state, a less-abstract layer of planning is necessary, or next state planning, to generate a near-term, 3-5 month plan to achieve a specific intermediate next state of the intervention. Looking at the bigger picture, next states divide a long-term intervention into a sequence of well-defined, short duration stepping stones. Figure 4 illustrates the relation between the overall intervention plan and the next states.

INSERT Figure 4: The next-states divide the overall intervention into short phases that present lower uncertainty range

At the beginning of each next-state phase, the intervention managers perform the following process (adapted here from (Kott 2007a) in a simplified form).

1. Assess the current situation and its relation to the intended end-state. Here, potentially, PMESII modeling tools help determine why the actions taken in the previous next state phase have not yielded the anticipated results. This may highlight an erroneous assumption. If there were major disparities between the model predictions and the real-world outcomes of the previous phase, this is the time to remodel specific phenomena.

2. Identify potentially feasible and desirable outcomes that describe favorable conditions of the next state (e.g., a national election).

3. Develop and analyze several alternative approaches to achieving the desired conditions of the next state. Form several planning teams and give each different assumption. The teams generate alternative strategies and analyze their effects. Here, PMESII modeling tools can play effective role: a planning team can use the tools to estimate and compare effects of alternative strategies, explore dependencies between various PMESII variables and their temporal dynamics, and test conceptual assumptions.

4. Select the best strategy, develop the next state plan, and perform risk and feasibility analyses.

5. Finalize and issue the action plan for execution.

Other authors emphasize the importance of periodic reassessment and replanning. For example, without offering a formal model or process for an interim-level replanning, Cuny and Hill (1999) provide recommendations for specific methods to assess periodically the state of famine in a region, and specific changes in famine response methods depending on the assessment.

The MPICE program (Dziedzic 2008) has developed a broad-ranging recommendation for gathering a variety of in-depth PMESII metrics that can be useful in assessing the nature of the current next state. With a set of standardized methodologies and tools for such assessments, PMESII modeling techniques will acquire a reliable baseline of data.

----------- // ------------

The key idea, in both of our examples, is a continuous cycle of anticipations and actions; in each cycle computational estimates of effects help intervention managers determine appropriate actions, and then assessments of real-world outcomes guide the next increment of computational estimates. With a proper methodology, PMESII modeling tools can offer valuable insights and encourage learning, even if they will never produce fully accurate estimates useable in a customary, strictly predictive manner.

Having flourished only within the last decade, PMESII modeling approaches exhibit the limitations of a very young discipline. Yet the trend is unmistakable: these approaches are maturing, gaining popularity, and becoming indispensible tools of analysts, planners, and decision-makers in government and business.

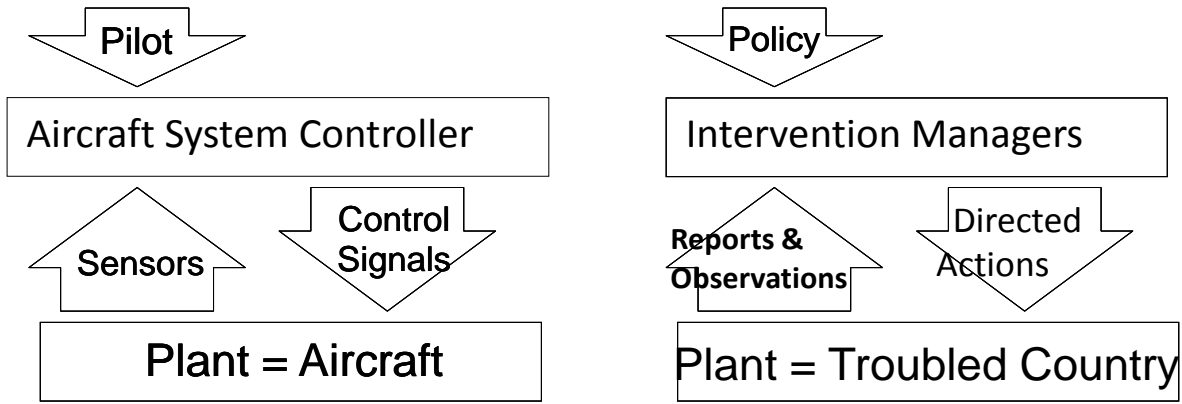

Figure 1: Variations on Control Theory

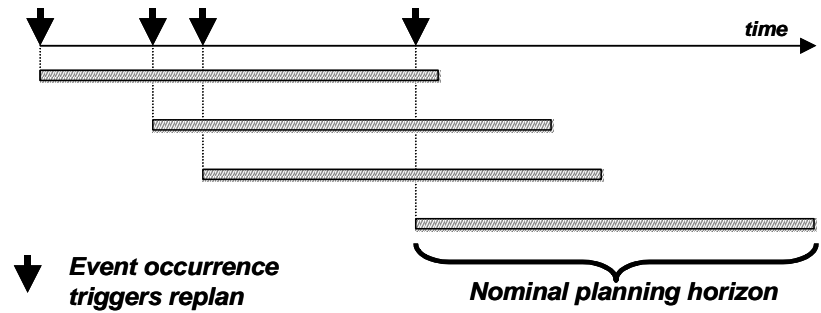

Figure 2 – An Approach to Replanning

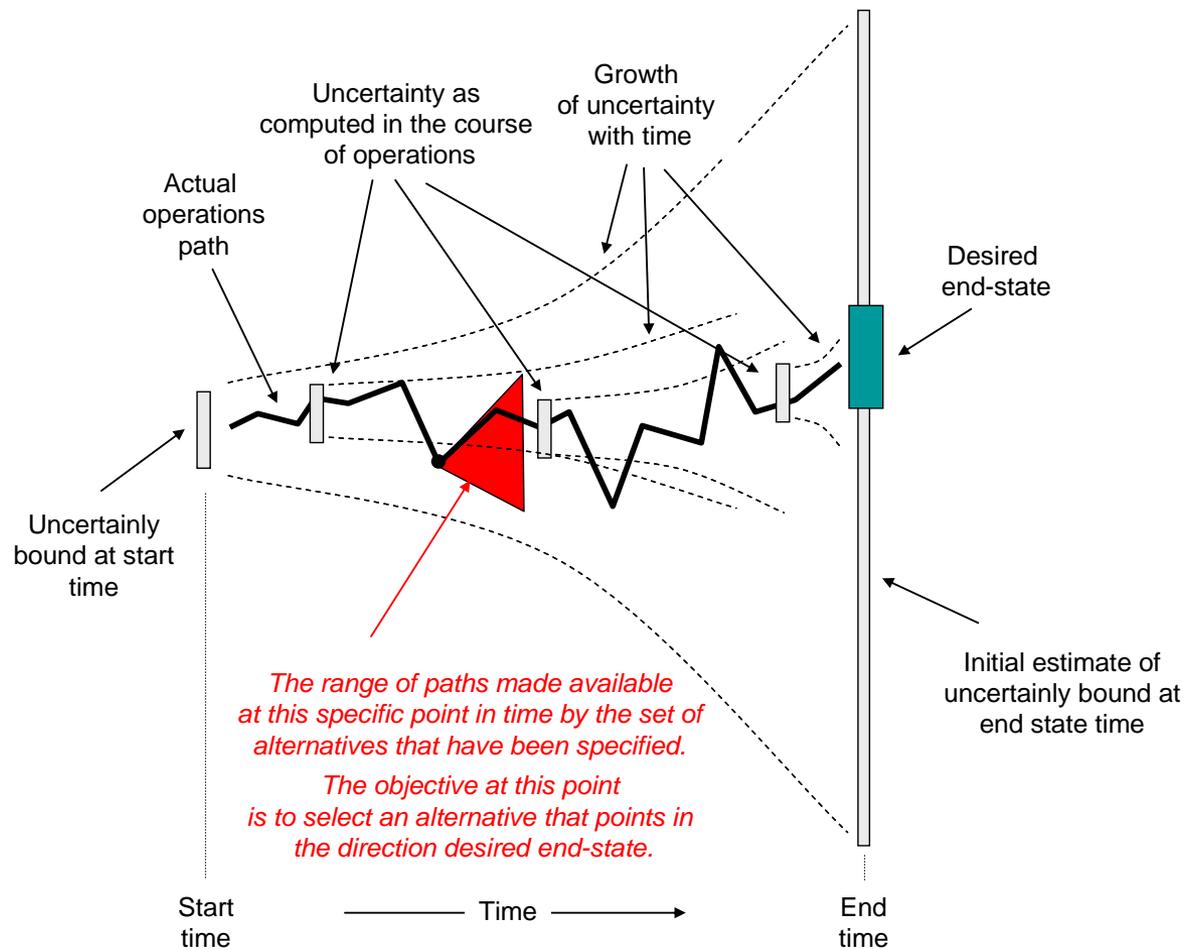

Figure 3: Planning Under Uncertainty

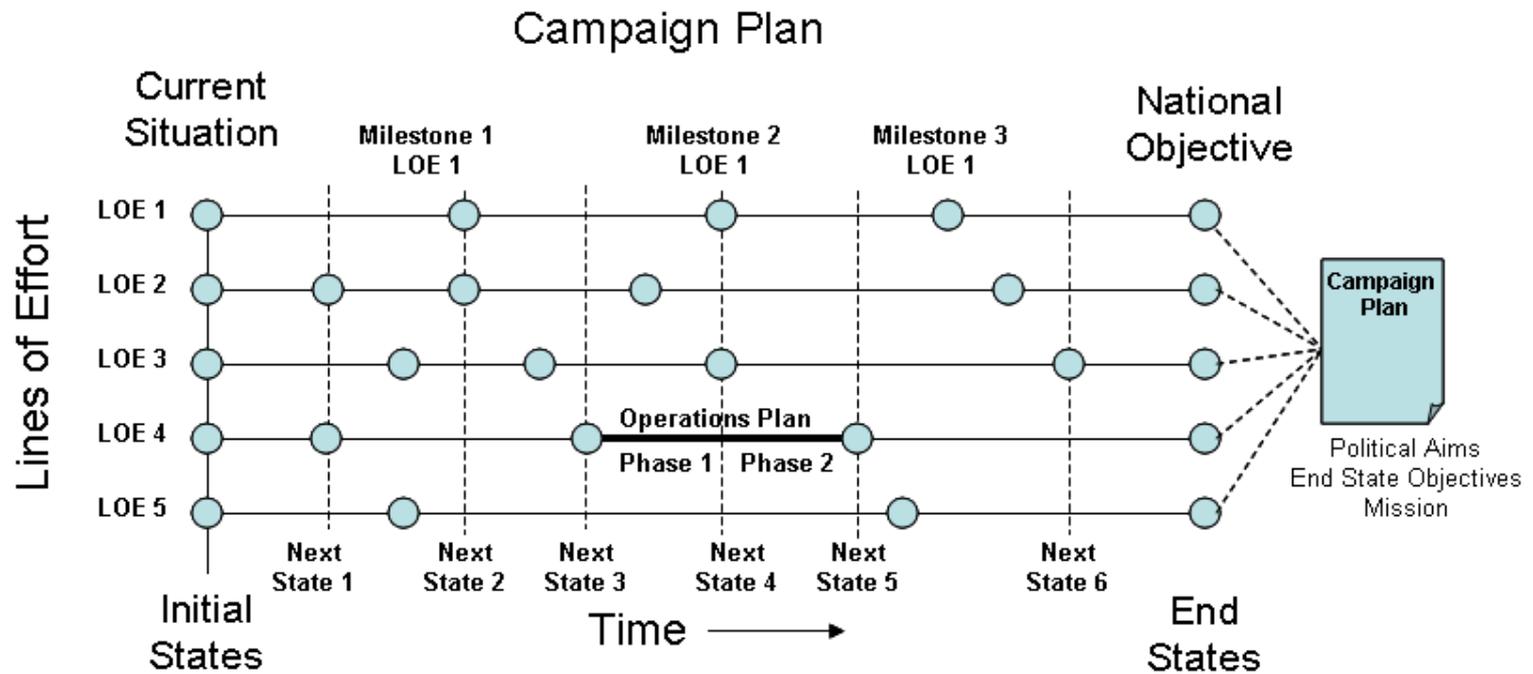

Figure 4: A Construct for Next State Planning